\documentclass[a4paper,fleqn]{cas-sc}

\usepackage[authoryear]{natbib}
\usepackage{caption}
\usepackage{subcaption}
\usepackage[utf8]{inputenc}

\def\tsc#1{\csdef{#1}{\textsc{\lowercase{#1}}\xspace}}
\tsc{WGM}
\tsc{QE}
\tsc{EP}
\tsc{PMS}
\tsc{BEC}
\tsc{DE}
\begin{document}
\let\WriteBookmarks\relax
\def\floatpagepagefraction{1}
\def\textpagefraction{.001}

\shorttitle{Enhancing Quranic Learning: A Multimodal Deep Learning Approach for Arabic Phoneme Recognition}

\shortauthors{Selim et~al.}

\title [mode = title]{Enhancing Quranic Learning: A Multimodal Deep Learning Approach for Arabic Phoneme Recognition}                      
%\tnotemark[1,2]

%\tnotetext[1]{This document is the results of the research
%   project funded by the National Science Foundation.}

%\tnotetext[2]{The second title footnote which is a longer text matter
%   to fill through the whole text width and overflow into
%   another line in the footnotes area of the first page.}

% Authors

% ---------- AUTHORS ----------
\author[a]{Ayhan K\"u\c{c}\"ukmanisa}
\ead{ayhan.kucukmanisa@kocaeli.edu.tr}

\author[b]{Derya Gelmez}
\ead{derya.glmz02@gmail.com}

\author[b,c]{\c{S}\"ukr\"u Selim \c{C}al\i k}
\ead{sselimcalik1@gmail.com}

\author[b]{Zeynep Hilal Kilimci}
\ead{zeynep.kilimci@kocaeli.edu.tr}

% ---------- CORRESPONDING NOTE ----------
\cortext[cor1]{Corresponding author: Ayhan K\"uç\"ukmanisa}

% ---------- ADDRESSES ----------
\address[a]{Department of Electronics and Communication Engineering, Kocaeli University, 41001, Kocaeli, Turkey}
\address[b]{Department of Information Systems Engineering, Kocaeli University, 41001, Kocaeli, Turkey}
\address[c]{Maviay Consultancy Company, Kocaeli University Technopark, 41275, Kocaeli, Turkey}


\begin{abstract}
Recent advances in multimodal deep learning have greatly enhanced the capability of systems for speech analysis and pronunciation assessment. Accurate pronunciation detection remains a key challenge in Arabic, particularly in the context of Qur’anic recitation, where subtle phonetic differences can alter meaning. Addressing this challenge, the present study proposes a transformer-based multimodal framework for Arabic phoneme mispronunciation detection that combines acoustic and textual representations to achieve higher precision and robustness. The framework integrates UniSpeech-derived acoustic embeddings with BERT-based textual embeddings extracted from Whisper transcriptions, creating a unified representation that captures both phonetic detail and linguistic context. To determine the most effective integration strategy, early, intermediate, and late fusion methods were implemented and evaluated on two datasets containing 29 Arabic phonemes, including eight hafiz sounds, articulated by 11 native speakers. Additional speech samples collected from publicly available YouTube recordings were incorporated to enhance data diversity and generalization. Model performance was assessed using standard evaluation metrics—accuracy, precision, recall, and F1-score—allowing a detailed comparison of the fusion strategies. Experimental findings show that the UniSpeech–BERT multimodal configuration provides strong results and that fusion-based transformer architectures are effective for phoneme-level mispronunciation detection. The study contributes to the development of intelligent, speaker-independent, and multimodal Computer-Aided Language Learning (CALL) systems, offering a practical step toward technology-supported Qur’anic pronunciation training and broader speech-based educational applications.
\end{abstract}

% \includegraphics{figs/grabs.pdf}

%\item To expedite future investigations, a proprietary dataset comprising Arabic letters is curated, incorporating the expertise of hafizes.
%\item Superior accuracy is achieved by audio-based transformer models in comparison to the current state-of-the-art studies.
%\item The proposed framework is designed to be speaker-independent, enabling it to operate effectively for any user without the requirement for individual enrollment.

% Keywords
\begin{keywords}

computer aided language learning \sep Arabic pronunciation detection \sep audio transformers \sep Whisper \sep UniSpeech \sep BERT \sep Multimodal

\end{keywords}

\maketitle

\section{Introduction}

The Holy Qur’an, revered as the divine scripture of the Islamic world, serves not only as a spiritual text but also as a central cultural, social, and educational reference. For millions of believers, the Qur’an is regarded as the literal Word of God, guiding every aspect of life. In this context, accurate recitation is essential, as it forms the foundation of both individual worship and collective ritual practice. Preserving the phonetic and semantic integrity of the original Arabic text is vital to conveying its meaning correctly. Arabic pronunciation, particularly in Qur’anic recitation, requires high precision: even a single mispronunciation can alter meaning or lessen the Qur’an’s aesthetic and spiritual impact. Therefore, developing automated tools that assist learners in achieving correct recitation is crucial for improving the quality of religious education and strengthening individuals’ engagement with the Qur’an, especially in communities where Arabic is not a native language \citep{elkheir2025quranic}.

Non-native learners of Arabic often face difficulties in mastering Qur’anic pronunciation. The phonetic and phonological structure of Arabic—including the \textit{makhraj} (points of articulation) and \textit{sifat} (articulatory characteristics) of letters—requires careful attention. Errors in pronouncing even a single sound may cause changes in meaning or obscure the intended message. Traditionally, identifying and correcting such mispronunciations has depended on qualified instructors, which limits access to effective training, particularly in regions with scarce educational resources. Automatic pronunciation assessment systems provide a promising alternative. By analyzing speech input, detecting pronunciation errors, and offering immediate feedback, they allow learners to practice independently while ensuring accessibility and consistency in evaluation \citep{algabri2022mdd,bahi2024automatic}.

In recent years, deep learning techniques have achieved significant progress in both natural language processing (NLP) and speech analysis. Among these methods, transformer-based architectures have gained prominence due to their ability to model long-term dependencies and capture linguistic subtleties that traditional approaches often miss. In speech and pronunciation tasks, such models can recognize fine acoustic differences with high accuracy \citep{vaswani2017attention}. Furthermore, multimodal learning, which integrates multiple data types such as audio, text, and visual inputs, enables a more comprehensive representation of speech phenomena. For example, in Arabic letter recognition, multimodal methods can combine the audio signal with corresponding textual or phonetic information—and, in some cases, visual cues such as lip movements—to enhance robustness and reduce ambiguity \citep{zhu2023multimodal}.

Based on these developments, the present study introduces a multimodal transformer-based system for Arabic letter recognition and mispronunciation detection in Qur’anic recitation. The proposed approach examines early, middle, and late fusion strategies for integrating acoustic and textual features. Transformer models are used to extract relevant information from audio signals, which is then combined with complementary modalities to produce a unified representation. These fusion strategies are compared to determine which method most effectively supports pronunciation accuracy. In addition to error detection, the system aims to serve as a practical learning aid, providing feedback that helps learners improve their recitation quality.

The main contributions of this study are summarized below:

\begin{itemize}
    \item A multimodal framework is developed for Arabic phoneme recognition in the Qur’anic context, systematically comparing early, mid, and late fusion strategies to clarify trade-offs and inform future research.
    \item The proposed system offers a practical and accessible solution for non-native Arabic speakers, supporting individualized learning and improving access to quality religious education.
    \item The study demonstrates the applicability of deep learning and transformer models in pronunciation assessment by adapting these methods to the domain of religious education.
    \item The research illustrates how technological advances can be effectively aligned with cultural and religious practices, providing a foundation for future interdisciplinary work.
\end{itemize}

The remainder of this paper is organized as follows. Section~\ref{sec2} reviews related literature on Qur’anic pronunciation challenges and previous research on automatic pronunciation evaluation for Arabic. Section~\ref{sec3} presents the proposed multimodal fusion framework, describing the early, mid, and late fusion strategies in detail. Section~\ref{sec4} outlines the dataset and experimental methodology, explaining how transformer-based architectures were applied for feature extraction and fusion. Section~\ref{sec5} reports and discusses the experimental results, and Section~\ref{sec6} concludes the paper with the main findings and directions for future work.

\section{Literature Review}\label{sec2}

This section reviews previous research on mispronunciation detection across different languages, with a particular focus on studies related to Arabic phonemes and Qur’anic recitation.

Early efforts in Arabic mispronunciation detection primarily relied on deep convolutional neural networks (CNNs). In \citep{nazir2019mispronunciation}, two CNN-based approaches were introduced: a CNN-feature extraction method and a transfer learning-based method. In the first approach, features extracted from intermediate CNN layers (layers 4–7) were classified using KNN, SVM, and neural network classifiers, while the second approach adapted a pre-trained CNN through transfer learning. The transfer learning-based method achieved the best performance, reaching an accuracy of 92.2\%, compared with 82\% for traditional hand-crafted features and 91.7\% for CNN-based features. Building on these results, \citep{akhtar2020improving} employed AlexNet to extract deep features from layers 6–8 and trained classifiers such as KNN, SVM, and Random Forest. With the inclusion of transfer learning and feature selection, the approach achieved an average accuracy of 93.2\%, highlighting the benefit of deeper and more discriminative CNN representations.

End-to-end (E2E) automatic speech recognition (ASR) architectures have also been explored for mispronunciation detection. The study by \citep{lo2020effective} introduced a CTC–Attention hybrid model that unifies alignment and recognition within a single framework, eliminating the need for phoneme-level forced alignment. Applied to Mandarin speech data, this approach demonstrated improved accuracy and simplified model design compared with traditional DNN–HMM systems.

Several studies have specifically targeted Classical Arabic pronunciation. In \citep{asif2021approach}, a deep neural network was trained on a newly collected dataset to recognize short vowels, achieving an accuracy of 95.77\%. Similarly, \citep{farooq2021mispronunciation} focused on detecting errors in the articulation points of Arabic letters using RASTA-PLP feature extraction combined with an HMM classifier. Reported accuracies reached 85\%, 90\%, and 98\% for different experimental setups, indicating the feasibility of technology-assisted pronunciation evaluation for Arabic. Another related work, \citep{alqadheeb2021correct}, collected 2,892 audio samples covering 84 short-vowel classes and employed a sequential CNN on 312 phonemes from the Arabic alphabet. The system achieved 100\% testing accuracy and a loss of 0.27, demonstrating the capability of CNN-based methods for precise Arabic phoneme classification.

For non-native Arabic learners, \citep{algabri2022mispronunciation} developed a comprehensive Computer-Assisted Pronunciation Training (CAPT) system capable of identifying mispronounced phonemes and their corresponding articulation features (AFs). The problem was formulated as a multi-label recognition task, and an additional speech corpus was synthesized using TTS technology. The best-performing model achieved a phoneme error rate (PER) of 3.83\%, an F1-score of 70.53\% for mispronunciation detection, and a detection error rate (DER) of 2.6\% for AF identification, outperforming previous end-to-end systems.

Beyond Arabic, recent studies have explored advanced neural architectures for multilingual pronunciation modeling. PeppaNet \citep{yan2023peppanet}, for example, proposed a unified end-to-end neural model that jointly performs alignment and dictation, integrating phonetic and phonological cues through a selective gating mechanism. Experiments on the L2-ARCTIC benchmark dataset showed notable improvements over earlier state-of-the-art systems.

In the specific context of Qur’anic recitation, \citep{harere2023mispronunciation} applied LSTM models with MFCC features to detect sequential pronunciation errors corresponding to Tajweed rules. Using the QDAT dataset, the model achieved accuracies of 96\%, 95\%, and 96\% for Separate Stretching, Tight Noon, and Hide rules, respectively, surpassing conventional machine learning baselines.

A recent study \citep{calik2023ensemble} proposed an ensemble-based approach for detecting mispronunciations of Arabic phonemes within a computer-assisted language learning (CALL) framework. The study introduced an ensemble model that integrates multiple conventional machine learning algorithms to identify mispronunciations and provide corrective feedback for Arabic pronunciation learning. To the best of the authors’ knowledge, this work represents the first comprehensive attempt to apply ensemble learning techniques to Arabic phoneme mispronunciation detection. In the experiments, mel-frequency cepstral coefficients (MFCC) and Mel-spectrogram features were extracted and combined with various classifiers to evaluate their impact on performance. The dataset consisted of recordings of 29 Arabic letters, including eight hafiz sounds, produced by 11 speakers, and was augmented through noise addition, time shifting, stretching, and pitch modification. Experimental results showed that the ensemble voting classifier using Mel-spectrogram features achieved the best performance, with an accuracy of 95.9\%, demonstrating the potential of ensemble learning for improving Arabic pronunciation assessment systems.

More recent work has shifted toward transformer-based and audio-oriented architectures. \citep{calik2024novel} compared SEW, HuBERT, Wav2Vec, and UniSpeech models for Arabic phoneme mispronunciation detection. Experiments conducted on two datasets covering 29 Arabic phonemes indicated that UniSpeech achieved the highest accuracy, while the study also discussed potential areas for improvement. Similarly, \citep{alrashoudi2025improving} developed a transformer-based approach for detecting and classifying phoneme-level errors by type (insertion, deletion, substitution). Using a dataset containing native and non-native Arabic speakers, the system achieved detection and diagnosis accuracies of 91.3\% and 80.8\%, respectively, outperforming human evaluators in certain conditions. 

Addressing the challenge of data imbalance, \citep{lounis2025oneclass} proposed a one-class CNN-based framework trained exclusively on correctly pronounced samples, treating mispronunciations as outliers. Experiments on the Arabic Speech Mispronunciation Detection Dataset (ASMDD) yielded an accuracy of approximately 84\%, demonstrating the potential of unsupervised or semi-supervised strategies for this domain. Complementing these developments, \citep{haouhat2025arabic} conducted a comprehensive review of Arabic multimodal machine learning (MML) studies, identifying major datasets, applications, and challenges, and outlining directions for future research.

In contrast to prior works, which often focus solely on acoustic cues or limited subsets of Arabic phonemes, the present study introduces a multimodal transformer-based framework that integrates both acoustic and textual information for Arabic phoneme mispronunciation detection. Furthermore, a dedicated dataset of Arabic letters has been constructed to support model training and benchmarking. This approach aims to enhance detection accuracy and robustness while providing a foundation for future studies on pronunciation assessment and computer-assisted Qur’anic education.

\section{Multimodal Fusion}\label{sec3}

Multimodal fusion is an approach that combines information from multiple data sources to take advantage of the complementary strengths of each modality. This integration helps overcome the limitations of relying on a single modality, promotes data diversity, and enables the creation of richer and more representative feature spaces. Fusion strategies are generally grouped into three main categories: early, intermediate, and late fusion.

In the early fusion approach, features extracted from different modalities are directly combined and processed by a single model. This allows the model to learn interactions between modalities at the feature level but may also introduce challenges such as increased dimensionality and potential incompatibility between modality scales. In contrast, intermediate fusion processes each modality independently within separate networks and merges their feature representations at selected intermediate layers. This strategy helps retain modality-specific characteristics while allowing a more balanced information exchange. The late fusion approach involves training independent classifiers for each modality and combining their outputs to reach a final decision. Late fusion preserves the independence of modalities and reduces the negative influence of potential errors from any single modality on the overall system performance \cite{li2024multimodal}.

In the current study, multimodal fusion techniques are applied to improve mispronunciation detection by integrating complementary information from acoustic and textual modalities. UniSpeech embeddings \cite{wang2021UniSpeech} are employed to capture fine-grained phonetic and prosodic cues, while BERT embeddings \cite{devlin2019bert} derived from phoneme-level transcriptions generated by Whisper \cite{radford2022robust} encode semantic, syntactic, and contextual information. The integration of these complementary representations forms a unified embedding space that models both phonetic and linguistic dimensions of speech. This shared representation aligns pronunciation patterns with their linguistic context, which is critical for accurate mispronunciation detection. Detailed descriptions of the models used in this study are provided in the following subsections.

\subsection{Robust Speech Recognition (Whisper)}

The Whisper model \cite{radford2022robust} represents a recent advancement in automatic speech recognition (ASR), designed to transcribe speech across different languages and challenging acoustic environments. Whisper employs a transformer-based encoder–decoder architecture capable of modeling long-range dependencies and temporal relationships within audio sequences. By processing raw audio waveforms sampled at 16 kHz, the model effectively encodes phonetic and prosodic information, yielding accurate and contextually consistent transcriptions.

A key feature of Whisper is its robustness to diverse acoustic conditions and speaker variations. Trained on a large-scale multilingual and multitask dataset, the model generalizes well across languages, accents, and speech styles. Its training strategy combines supervised learning on labeled datasets with multitask objectives, enhancing adaptability and enabling reliable transcription performance even in noisy or non-standard speech. 

Whisper’s self-attention mechanisms capture contextual relationships across complete utterances, allowing the model to maintain coherence between phonemes and words. This makes it particularly suitable for downstream applications such as phoneme-level analysis, speech translation, and pronunciation assessment. In summary, Whisper provides a robust and versatile foundation for generating phoneme-level transcriptions that are used as textual input in the proposed multimodal framework.

\subsection{Unified Speech Representation Learning (UniSpeech)}

The UniSpeech model \cite{wang2021UniSpeech} is a transformer-based architecture developed to learn effective speech representations for a wide range of tasks, including automatic speech recognition, language identification, and speaker verification. By leveraging self-attention mechanisms, UniSpeech captures both local and global dependencies within audio signals, enabling the extraction of detailed acoustic features that reflect the hierarchical structure of speech.

A major strength of UniSpeech lies in its multilingual and multitask training design. Trained on large multilingual datasets, the model generalizes efficiently across languages and tasks, making it suitable for applications involving low-resource or cross-lingual speech processing. Furthermore, UniSpeech incorporates unsupervised and semi-supervised learning strategies that allow it to utilize vast amounts of unlabeled audio data, improving robustness against variability in accents, recording conditions, and speaking styles. The use of transfer learning and knowledge distillation further enhances its adaptability, enabling the transfer of learned knowledge from high-resource to low-resource domains.

Extensive benchmark evaluations have shown that UniSpeech achieves competitive results across various speech processing tasks. In this study, UniSpeech embeddings are used to represent the acoustic characteristics of Arabic phonemes, capturing subtle phonetic and prosodic features that are essential for accurate mispronunciation detection.

\subsection{Contextual Language Understanding (BERT)}

The BERT model \cite{devlin2019bert} is a transformer-based language representation model that produces deep contextual embeddings from textual data. Through its bidirectional self-attention mechanism, BERT captures both semantic meaning and syntactic structure, allowing it to interpret words and phonemes in relation to their surrounding context.

BERT is pretrained using masked language modeling and next sentence prediction objectives, enabling it to learn general-purpose language representations from large-scale unlabeled text corpora. When fine-tuned on phoneme-level transcriptions produced by Whisper, BERT generates context-aware textual embeddings that complement the acoustic representations obtained from UniSpeech. These embeddings encode linguistic correctness, semantic coherence, and phonetic consistency, supporting the identification of pronunciation errors within linguistic context.

Integrating BERT embeddings with UniSpeech acoustic features allows the proposed framework to jointly consider linguistic and phonetic dimensions of speech. This multimodal alignment enhances the system’s capacity to detect nuanced and context-dependent mispronunciations. Consequently, BERT serves as a reliable textual encoder within the proposed multimodal learning framework for Arabic phoneme mispronunciation detection.

\section{Proposed framework}\label{sec4}

\begin{figure}
  \centering
  \includegraphics[width=1.0\textwidth]{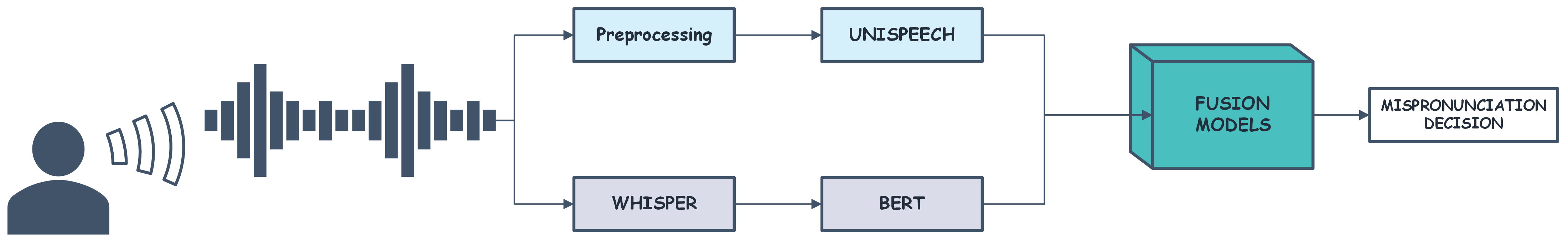}
  \caption{The block diagram of proposed method}
  \label{fig:block}
\end{figure}

This section presents the proposed multimodal framework designed to detect mispronunciations of Arabic phonemes by integrating acoustic and textual information. The framework combines state-of-the-art transformer models to capture both phonetic and linguistic cues, forming a unified representation that supports accurate pronunciation assessment. It consists of three main stages: data preprocessing and feature extraction, multimodal fusion of acoustic and textual embeddings, and classification of phoneme-level pronunciations. The following subsections describe the multimodal fusion strategy and the models employed in detail, including Whisper for transcription generation, UniSpeech for acoustic representation learning, and BERT for contextual language understanding. The block diagram of proposed method is illustrated in Fig. \ref{fig:block}.

\subsection{Dataset \& Preprocessing}
The dataset used in this study was adopted from the work of Çalık et al. \citep{calik2024novel}.
This dataset consists of audio recordings representing the pronunciation characteristics of the Arabic alphabet and was originally constructed using two main sources: direct recordings from Qur’anic reciters (Hafiz) and publicly available audio clips obtained from YouTube.
It contains a total of 1015 audio samples, including 232 directly recorded Arabic alphabet sounds and 783 distinct recordings collected from online sources, covering the full range of Arabic phonemes.

The raw audio data underwent a comprehensive preprocessing pipeline to ensure high performance during model training. The initial step of this process was to remove background noise from the recordings, thereby enhancing the audio quality. Subsequently, all audio files were standardized and unified into a common format. This standardization included resizing each audio clip to a duration of 4 seconds and resampling to a frequency of 16 kHz.

Following the completion of the data preprocessing, two distinct datasets (Dataset A and Dataset B) are used to evaluate the model's performance under different scenarios. Details of these dataset setups are given in Table \ref{tab:datasets}.

Two dataset configurations were designed to evaluate the model under different training–testing conditions. 
In \textit{Dataset A}, the 783 audio samples collected from YouTube were used for training, while the 232 recordings obtained from Qur’anic reciters were reserved for testing, enabling the assessment of generalization on expert pronunciations. 
In \textit{Dataset B}, 80\% of both YouTube and reciter recordings were used for training and the remaining 20\% for testing, allowing evaluation on a more homogeneous data distribution. 
These configurations provided complementary perspectives on model robustness and generalization capability.

\begin{table}[!htbp]
\centering
\caption{Dataset setups with sample and person distributions for YouTube and Hafiz sources.}
\label{tab:datasets}
\setlength{\tabcolsep}{5pt}
\renewcommand{\arraystretch}{1.2}
\begin{tabular}{ccccccccc}
\toprule
 & \multicolumn{4}{c}{Train} & \multicolumn{4}{c}{Test} \\
\cmidrule(lr){2-5} \cmidrule(lr){6-9}
 & \multicolumn{2}{c}{YouTube} & \multicolumn{2}{c}{Hafiz} 
 & \multicolumn{2}{c}{YouTube} & \multicolumn{2}{c}{Hafiz} \\
\cmidrule(lr){2-3} \cmidrule(lr){4-5} \cmidrule(lr){6-7} \cmidrule(lr){8-9}
Dataset & Sample & Person & Sample & Person & Sample & Person & Sample & Person \\
\midrule
A & 783 & 35 & 0 & 0 & 0 & 0 & 232 & 11 \\
B & 626 & 28 & 186 & 9 & 157 & 7 & 46 & 2 \\
\bottomrule
\end{tabular}
\end{table}

\subsection{Proposed Method}

In this study, a multimodal framework integrating both acoustic and textual modalities is developed for the detection of mispronunciations in Arabic phonemes. 
The proposed methodology consists of several stages, including feature extraction, model design, cross-validation training, fusion strategy implementation, and performance evaluation.

For the acoustic stream, the feature encoders of the proposed Transformer models operate over a receptive field covering approximately 400 samples (about 25 ms) of audio. 
Since the encoding process is conducted using signals sampled at 16 kHz, all input audio recordings are resampled to 16 kHz during preprocessing. 
The audio inputs are then padded or trimmed to a fixed duration of 4 seconds before being fed into the acoustic encoder.

A pre-trained UniSpeech Transformer model is employed to extract acoustic embeddings. 
For the textual modality, all audio recordings are processed using the Whisper model, which provides phoneme-level transcriptions for each utterance. 
Each audio sample is paired with its corresponding phoneme sequence and subsequently transformed into high-dimensional embeddings using a pre-trained multilingual BERT model.

The pre-trained weights of Transformer models trained on Arabic speech data are adopted as the initial weights in the proposed framework to improve convergence. Subsequently, each proposed model is fine-tuned using Qur’anic recitation data, allowing it to capture pronunciation variations and phonetic characteristics unique to Qur’anic Arabic. To prevent overfitting and to reliably assess the generalizability of the models, a five-fold cross-validation procedure is applied to the training data of both datasets (A and B). Each training set is divided into five equal subsets, where four subsets are used for training and one subset is used for validation in each iteration. This process ensures that each subset serves once as a validation set, resulting in a total of five independent training–validation cycles. After completing all cross-validation cycles, the model weights that achieve the best validation performance are selected, and the resulting models are evaluated on the test set, which is kept completely unseen during training. 
This strategy provides an objective estimation of the models' true generalization capability.

Three fusion strategies were implemented to integrate the complementary information provided by the acoustic and textual modalities. The architecture of the proposed deep learning-based fusion models is depicted in Figure \ref{fig:placeholder}.

Early Fusion: In this approach, acoustic embeddings obtained from the UniSpeech encoder and textual embeddings extracted from the BERT model were concatenated immediately after the feature extraction stage. To ensure numerical compatibility between the two modalities, embeddings were normalized prior to concatenation. The fused vector was then processed by a classification network composed of multiple fully connected layers with ReLU activations and dropout regularization. This design enabled the model to jointly capture acoustic and textual information from the initial stage of learning, directly modeling cross-modal dependencies within the classifier.

Intermediate Fusion: In this approach, each modality was first processed independently before integration. UniSpeech and BERT embeddings were passed through separate networks to reduce dimensionality and highlight salient, modality-specific features. The resulting transformed embeddings were subsequently concatenated and fed into a classifier network. This procedure allowed the model to first optimize each modality independently and then perform multimodal integration.

Late Fusion: In this approach, the unimodal AudioOnly and TextOnly models were trained independently to convergence. Hidden-layer embeddings from each model were extracted as modality-specific representations and passed through layers to reduce dimensionality and emphasize the most salient features. The reduced representations were concatenated and processed by a classification network consisting of fully connected layers with ReLU activations and dropout regularization. During this stage, the pretrained encoder weights of UniSpeech and BERT were frozen, ensuring that only the fusion layers were updated and the individual encoders were not retrained.

All models were evaluated using accuracy, precision, recall, F1-score. All evaluations were conducted on the two distinct datasets (A and B).

This methodology enables the integration of self-supervised acoustic embeddings with phoneme-level textual embeddings through early, intermediate, and late fusion strategies, allowing robust and accurate identification of mispronunciations in Arabic phonemes under varying recording conditions.

\begin{figure}
    \centering

    \begin{subfigure}{0.30\linewidth}
        \centering
        \includegraphics[width=\linewidth]{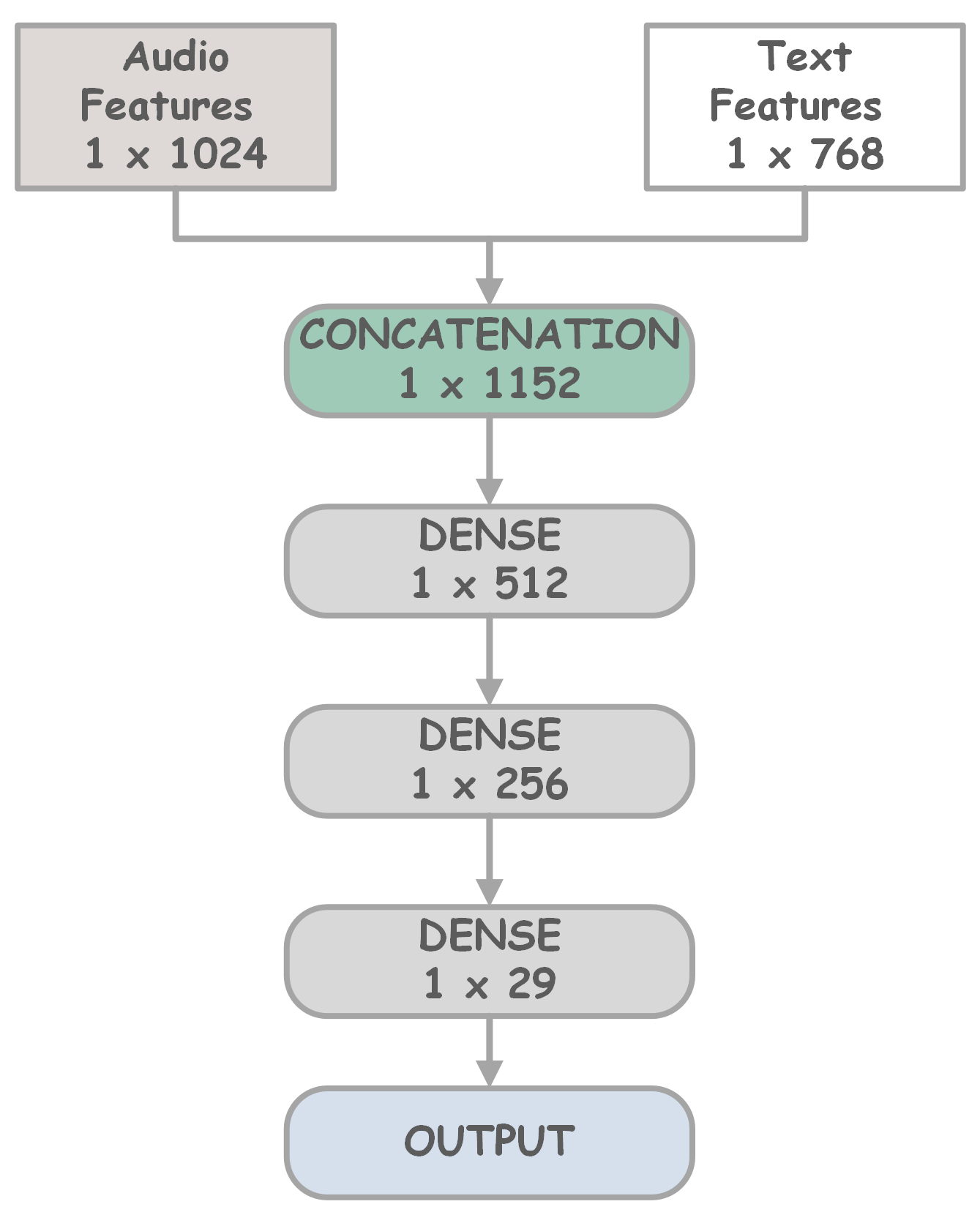}
        \caption{Early Fusion}
        \label{fig:fusion_early}
    \end{subfigure}
    \hfill
    \begin{subfigure}{0.33\linewidth}
        \centering
        \includegraphics[width=\linewidth]{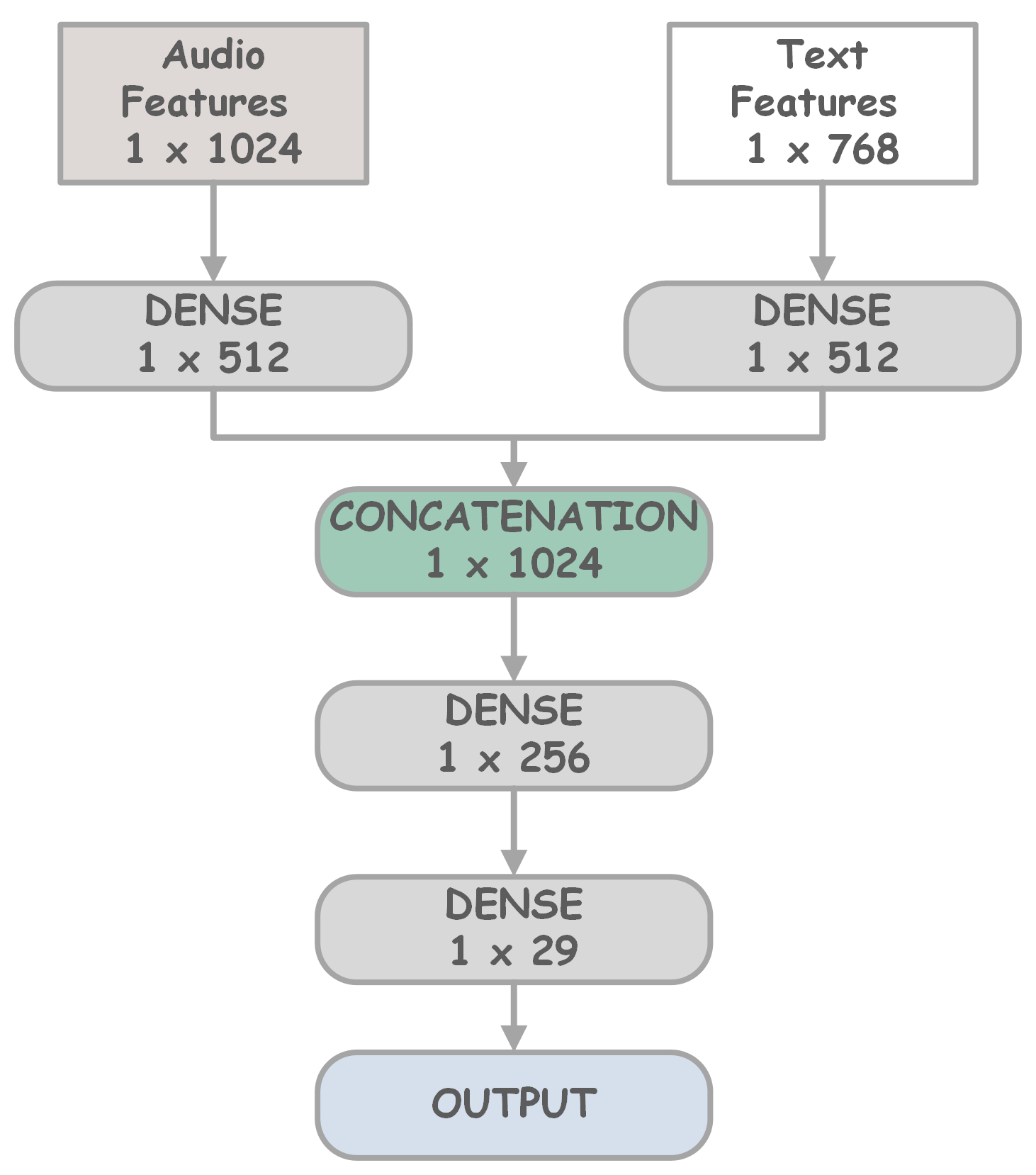}
        \caption{Intermediate Fusion}
        \label{fig:fusion_intermediate}
    \end{subfigure}
    \hfill
    \begin{subfigure}{0.33\linewidth}
        \centering
        \includegraphics[width=\linewidth]{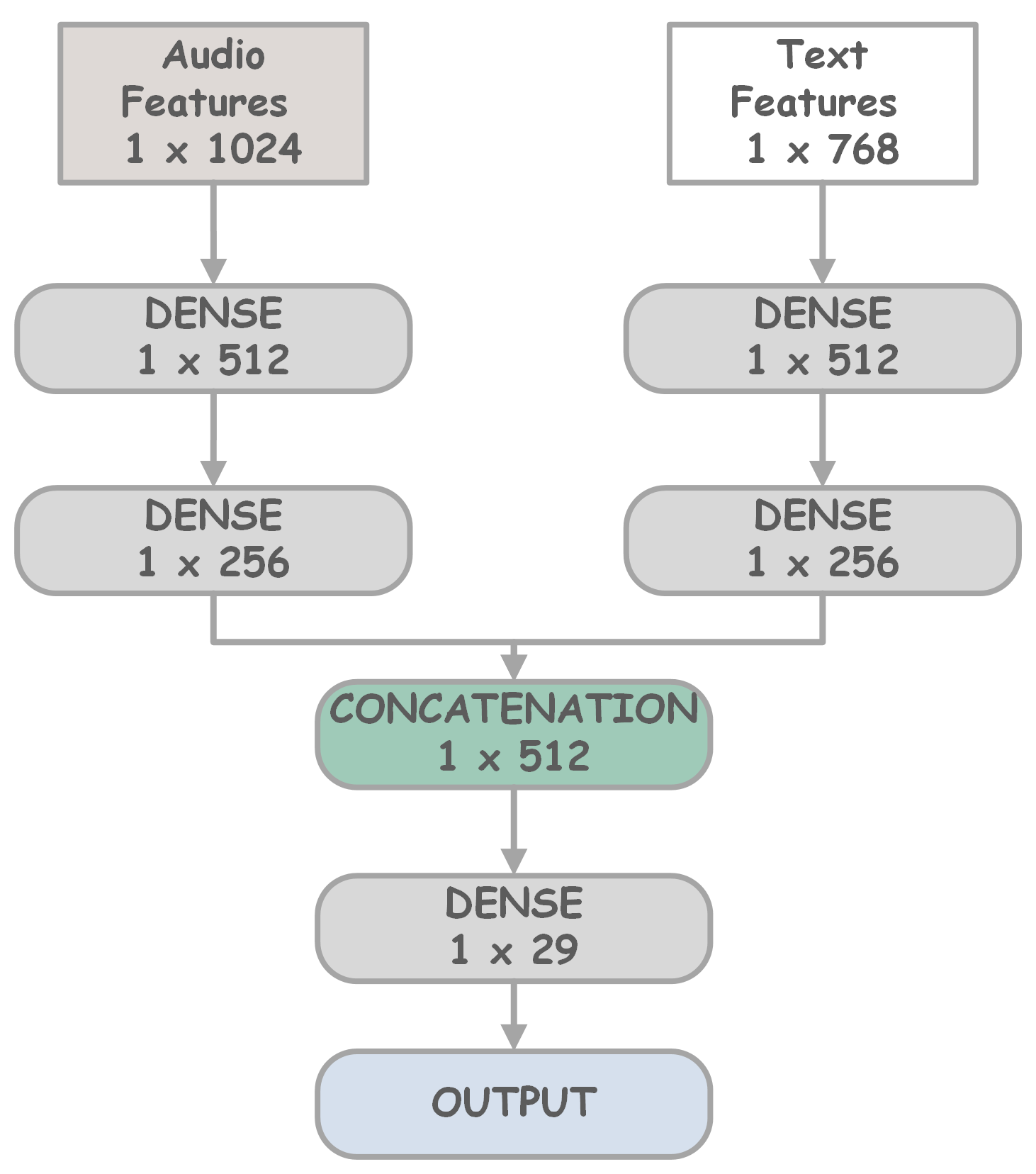}
        \caption{Late Fusion}
        \label{fig:fusion_late}
    \end{subfigure}

    \caption{The proposed deep learning-based fusion models (a) Early fusion, (b) Intermediate fusion, and (c) Late fusion.}
    \label{fig:fusion_models}
\end{figure}

\section{Experimental results}\label{sec5}

The evaluation of the proposed methods is quantified through the utilization of the formulas provided in equations (1), (2), (3), and (4). Within these equations, each class is designated as $ C_{x} $, where TP (True Positive) signifies accurate classification of audio belonging to the $ C_{x} $ class as $ C_{x} $ itself. FP (False Positive) pertains to all non-$ C_{x} $ samples erroneously classified as $ C_{x} $. TN (True Negative) encompasses all non-$ C_{x} $ samples correctly not classified as $ C_{x} $. FN (False Negative) encapsulates instances where samples from $ C_{x} $ are inaccurately not classified as $ C_{x} $.

\label{equations}
\begin{equation}
Accuracy = \frac{TP+TN}{TP+TN+FP+FN}
\end{equation}

\begin{equation}
Precision = \frac{TP}{TP+FP}
\end{equation}

\begin{equation}
Recall = \frac{TP}{TP+FN}
\end{equation}

\begin{equation}
F-measure = 2 \times \frac{Precision \times Recall}{Precision + Recall}
\end{equation}

The training parameters for the proposed multimodal models are kept consistent across all three fusion strategies. Each model is trained with a batch size of 8 and a learning rate of $3\times10^{-5}$, optimized using the AdamW optimizer. The maximum number of epochs is set to 30 for all fusion configurations.

To ensure fair hyperparameter selection, a grid search–based tuning procedure is conducted on the validation set, where candidate values for the learning rate, batch size, and dropout rate are systematically evaluated. The selected configuration represents the best-performing combination identified through this search process.

To prevent overfitting, an early stopping mechanism is applied, which automatically terminates training when the validation loss fails to improve for a predefined number of epochs. This approach ensures that the models converge at the point yielding the best generalization performance without unnecessary overtraining.

The results presented in Table~\ref{tab:results_fusion} indicate that multimodal fusion strategies applied to the UniSpeech + BERT framework provide strong classification performance on both datasets. For Dataset A, the Early and Intermediate Fusion strategies achieved identical top performance across all evaluation metrics (Accuracy: 0.966, F1-Score: 0.965), demonstrating that cross-modal interactions are equally effective whether performed at the initial feature level or within intermediate network layers. This suggests that both fusion points enable efficient exploitation of complementary acoustic and linguistic cues.

On Dataset B, the Intermediate Fusion approach delivered the best results (Accuracy: 0.985, F1-Score: 0.985), outperforming both Early and Late Fusion configurations. This indicates that aligning modality-specific features at a deeper layer of the network can yield superior representations when handling more diverse and challenging samples. Meanwhile, Early Fusion still maintains strong competitive performance, confirming its reliability as a robust and efficient baseline fusion strategy.

Conversely, Late Fusion consistently showed the lowest performance across both datasets, highlighting that combining independent decisions at the output stage may limit synergistic feature learning. Since high-level modality interactions occur only after classification, this approach fails to fully leverage cross-modal relationships embedded in the data.

Overall, these findings demonstrate that feature-level and intermediate-level fusion are more effective than decision-level integration, and that the optimal fusion position can vary depending on dataset complexity. The results support the conclusion that multimodal representation learning—rather than unimodal transformer processing—drives the substantial improvements observed in downstream classification performance.

Table~\ref{tab:results_SOTA} compares the performance of the proposed multimodal fusion model with the existing transformer-based state-of-the-art (SOTA) method. 

For Dataset A, the proposed multimodal fusion model achieves an accuracy of 0.966, outperforming the transformer-based SOTA baseline by +2.2 percentage points, alongside consistent improvements across Precision, Recall, and F1-Score. This demonstrates that jointly modeling acoustic and linguistic information at the feature level enriches the representational space and enhances discriminative capability beyond what a unimodal transformer can achieve.

The improvement becomes more evident for Dataset B, where the proposed approach reaches an accuracy of 0.985 versus 0.970 with the SOTA method, reflecting a +1.5 percentage-point increase. Again, all metrics show parallel gains, confirming that combining BERT’s contextual language understanding with UniSpeech’s speech-specific embeddings yields a more robust and generalizable classifier across varying data conditions.

A key observation is that this advancement is achieved without increasing the complexity of a single transformer model. Instead of scaling a unimodal transformer to deeper or more parameter-heavy configurations, the proposed approach distributes learning across two modality-adapted transformer encoders and integrates them through an effective fusion mechanism. This highlights that performance improvements stem from cross-modal synergy rather than model size or architectural depth.

Overall, the findings confirm that the proposed multimodal fusion strategy offers a more accurate and efficient alternative to purely transformer-based unimodal models.

\begin{table}[!t]
\centering
\caption{Evaluation results of proposed fusion models on both datasets}
\label{tab:results_fusion}
\begin{tabular}{cccccc}
\hline
\textbf{Dataset} & \textbf{Fusion Method} & \textbf{Accuracy} & \textbf{Precision} & \textbf{Recall} & \textbf{F1-Score} \\ \hline

A & Early         & \textbf{0.966} & \textbf{0.969} & \textbf{0.966} & \textbf{0.965} \\
A & Intermediate  & \textbf{0.966} & \textbf{0.969} & \textbf{0.966} & \textbf{0.965} \\
A & Late          & 0.957          & 0.959          & 0.957          & 0.957          \\ \hline

B & Early         & 0.970          & 0.974          & 0.970          & 0.970          \\
B & Intermediate  & \textbf{0.985} & \textbf{0.988} & \textbf{0.985} & \textbf{0.985} \\
B & Late          & 0.956          & 0.964          & 0.956          & 0.955          \\ \hline

\end{tabular}
\end{table}

\begin{table}[!t]
\centering
\caption{Performance comparison with the state-of-the-art (SOTA)}
\label{tab:results_SOTA}
\begin{tabular}{cccccc}
\hline
\textbf{Dataset} & \textbf{Method} & \textbf{Accuracy} & \textbf{Precision} & \textbf{Recall} & \textbf{F1-Score} \\ \hline

A & \citep{calik2024novel}        & 0.944 & 0.950 & 0.944 & 0.943 \\
A & Proposed Method    & \textbf{0.966} & \textbf{0.969} & \textbf{0.966} & \textbf{0.965} \\ \hline

B & \citep{calik2024novel}        & 0.970 & 0.966 & 0.970 & 0.962 \\
B & Proposed Method    & \textbf{0.985} & \textbf{0.988} & \textbf{0.985} & \textbf{0.985} \\ \hline

\end{tabular}
\end{table}

\section{Discussion}

The experimental results show that the effectiveness of the proposed multimodal fusion approach, which integrates UniSpeech-based acoustic representations and BERT-based textual embeddings to enhance speech–text classification performance. Across both datasets, the fusion model outperformed the transformer-based state-of-the-art (SOTA) method. This indicates that combining modality-specific representations provides complementary and more distinctive information than relying on a single transformer architecture.

A key finding of this study is the clear benefit of cross-modal feature interaction. While transformer models such as UniSpeech or BERT individually capture high-quality representations within their respective domains, their unimodal nature limits the contextual richness required for robust decision-making in tasks where both speech and linguistic context play a role. The fusion-based model mitigates this limitation by bringing together prosodic, spectral, and semantic information, resulting in more comprehensive and expressive latent feature spaces. This synergy was particularly evident in the substantial gains seen in accuracy, precision, recall, and F1-score, especially on Dataset B, where the model achieved up to a 1.5 percentage-point improvement over the SOTA baseline.

Another notable aspect is that these improvements were achieved without increasing the architectural depth or computational burden of a single, monolithic transformer. Instead of scaling the size of the model, an approach commonly adopted to boost performance in transformer-based systems, the proposed solution distributes learning across two domain-optimized encoders and leverages a fusion network to integrate knowledge. This shows that, for multimodal classification, strategic fusion design can be more impactful and efficient than increasing model complexity, offering both performance and computational advantages. Such findings align with emerging trends in multimodal learning, which emphasize the importance of leveraging heterogeneous cues rather than relying solely on model scaling.

In summary, the findings validate that multimodal integration of speech and text features provides a stronger foundation for classification tasks compared to unimodal transformer models, offering enhanced accuracy, improved robustness, and better generalization without added architectural complexity. The insights derived from this study can guide the development of next-generation multimodal systems and encourage a shift from model-centric scaling toward modality-centric fusion strategies.

\subsection{Research Limitations}
Despite its contributions, several limitations should be acknowledged to contextualize the scope of the study.
First, the datasets used for training and evaluation—although enriched with diverse samples including YouTube recordings—remain relatively limited in scale and dialectal coverage. Arabic is characterized by significant regional and sociolinguistic variability; thus, models trained on a restricted subset of dialects may not fully generalize to all pronunciations or accents. Future large-scale data collection efforts are necessary to ensure broader representativeness.

Second, the speech data primarily consist of curated and relatively clean audio recordings, which may not accurately reflect real-world acoustic conditions. Environmental noise, microphone variability, and spontaneous speech phenomena such as hesitations and co-articulations were not extensively represented in the dataset, limiting the framework’s robustness in uncontrolled environments.

Third, the study focused exclusively on transformer-based fusion architectures. While these models have shown strong performance, other multi modal learning paradigms—such as convolutional-recurrent hybrids, cross-attention networks, or graph-based fusion methods—were not explored. Additionally, model interpretability was beyond the present study’s scope; a deeper investigation into which modality or feature subset contributes most to mispronunciation detection could yield valuable insights for linguistic and pedagogical applications.

Finally, although the evaluation metrics (accuracy, precision, recall, and F1-score) provided quantitative insight into model performance, subjective human evaluation or expert linguistic assessment was not included. Such qualitative validation could further substantiate the model’s pedagogical effectiveness in CALL systems.

\subsection{Potential Future Research}

Building on these findings, several promising avenues exist for future exploration. First, expanding the dataset to include additional speakers, genders, dialects, and recitation styles would enhance model generalization and enable cross-dialectal robustness. Collecting spontaneous, noisy, and emotionally varied speech samples would also allow more realistic testing of model resilience. Second, integrating real-time pronunciation feedback mechanisms into the multimodal framework could transform it into an interactive CALL tool. Embedding the model within mobile or web-based platforms would allow learners to receive immediate, context-aware feedback on pronunciation errors, potentially accelerating learning outcomes and improving retention. Third, future studies may investigate adaptive and personalized learning mechanisms using reinforcement learning or speaker embedding techniques, enabling the system to tailor feedback based on the learner’s proficiency and progress. Such approaches could bridge the gap between automatic assessment and individualized pedagogy. Fourth, exploring cross-lingual and multilingual multimodal fusion could extend this methodology to other morphologically rich or phonemically intricate languages, such as Persian, Urdu, or Turkish. Comparative studies across languages may also reveal universal patterns in multimodal phoneme learning.

Moreover, incorporating prosodic, articulatory, and visual cues—for instance, through lip-reading data or spectrogram-based articulatory feature modeling—could provide a more holistic understanding of pronunciation and articulation. Finally, a potential future direction involves combining multimodal transformers with self-supervised pretraining on large multilingual corpora, enhancing both transferability and efficiency in low-resource contexts. Collectively, these directions underscore that the present research not only advances the state of Arabic mispronunciation detection but also establishes a generalizable paradigm for multimodal speech assessment, bridging the gap between linguistic theory, machine learning, and educational technology.\\

\noindent\textbf{Declaration of Generative AI and AI-Assisted Technologies in the Writing Process}

\noindent During the preparation of this work the authors used ChatGPT tool in order to improve language and readability. After using this tool, the authors reviewed and edited the content as needed and takes full responsibility for the content of the publication.


\bibliographystyle{cas-model2-names}

\bibliography{cas-refs}


%Author biography without author photo.
%Author biography. Author biography. Author biography.
%Author biography. Author biography. Author biography.
%Author biography. Author biography. Author biography.
%Author biography. Author biography. Author biography.
%Author biography. Author biography. Author biography.
%Author biography. Author biography. Author biography.
%Author biography. Author biography. Author biography.
%Author biography. Author biography. Author biography.
%Author biography. Author biography. Author biography.

%\bio{figs/pic1}
%Author biography with author photo.
%Author biography. Author biography. Author biography.
%Author biography. Author biography. Author biography.
%Author biography. Author biography. Author biography.
%Author biography. Author biography. Author biography.
%Author biography. Author biography. Author biography.
%Author biography. Author biography. Author biography.
%Author biography. Author biography. Author biography.
%Author biography. Author biography. Author biography.
%Author biography. Author biography. Author biography.

%\bio{figs/pic1}
%Author biography with author photo.
%Author biography. Author biography. Author biography.
%Author biography. Author biography. Author biography.
%Author biography. Author biography. Author biography.
%Author biography. Author biography. Author biography.
%\endbio %

\end{document}